# Assessing Subgroup Heterogeneity in Carryover Effects: A Simulation Study and an Observational Alzheimer Disease Treatment-Transition Application

Zimeng Ren[1], Kun Huang[1], Yongming Qu[2], Wanzhu Tu[1], Xiaoqing Huang[1,*]

[1] Department of Biostatistics, Indiana University School of Medicine, Indianapolis, IN, 46202, USA.

[2] Department of Global Statistical Sciences and Advanced Analytics, Eli Lilly and Company, Indianapolis, IN. 46225. USA.

[*]Corresponding author: huanxi@iu.edu

## Abstract

Subgroup heterogeneity in carryover effects is not routinely evaluated in 2×2 crossover studies, and its inferential consequences remain insufficiently characterized. When the scientific question concerns subgroup-specific carryover, omission of the corresponding interaction causes the fitted model to target a collapsed population-average contrast rather than the subgroup difference of interest. We used simulation to evaluate the finite-sample bias, confidence-interval coverage, Type I error, and power of carryover-by-subgroup interaction inference under an independence-based analysis and REML-based MMRM analyses with CS and UN covariance specifications. Reduced mean models were included as population-average benchmarks. We also applied the models to observational Alzheimer disease treatment-transition sequences from NACC, where carryover was defined operationally as a model-based association with the preceding regimen rather than an identified pharmacologic persistence effect. Full models directly estimated subgroup-specific carryover heterogeneity, whereas reduced models estimated a collapsed carryover contrast that did not characterize this heterogeneity. Under the full mean model, point-estimate bias was small, whereas covariance misspecification affected empirical confidence-interval coverage and Type I error. Reliable inference on subgroup-specific carryover effects requires a mean model that explicitly represents the heterogeneity of interest and a covariance structure appropriate for the repeated-measures data. Failure to address either component can obscure subgroup differences or compromise statistical inference. These results provide practical guidance for evaluating heterogeneous carryover effects in crossover and treatment-transition studies.

## 1 | Introduction

Crossover designs are widely used in clinical research when within-subject comparisons are scientifically meaningful, the outcome is sufficiently stable or reversible, and carryover effects from prior treatment can be adequately addressed [1, 2]. In a two-sequence, two-period crossover trial, participants are randomly assigned to receive treatment A followed by treatment B or treatment B followed by treatment A. Because participants serve as their own control, crossover designs may improve statistical efficiency and power relative to parallel-group designs with the same sample size, provided that the assumptions underlying within-subject comparison are satisfied. However, period and sequence effects, as well as residual carryover from prior treatment, may complicate estimation and interpretation of treatment effects in practice [2]. Although washout periods are intended to mitigate carryover, residual effects may persist when treatment effects are slow to dissipate or when outcomes reflect behavioral, symptomatic, or other processes that are not rapidly reversible [3, 4].

An additional challenge arises when treatment or carryover effects differ across clinically relevant patient subgroups. Subgroups defined by demographic characteristics, baseline disease severity, biomarker status, disease stage, or other clinical characteristics may differ in their responses to treatment [5, 6]. Formal assessment of such heterogeneity requires specification and testing of the relevant treatment-by-subgroup interactions rather than separate or post hoc comparisons of subgroup-specific estimates [7, 8]. This principle is increasingly emphasized in clinical research because population-average effects can obscure clinically meaningful differences across patient groups, with implications for treatment selection, benefit-risk assessment, and generalizability [9-13]. Despite this increasing clinical and methodological emphasis, rigorous implementation remains incomplete. For example, the Consolidated Standards of Reporting Trials (CONSORT) 2025 guidance [13] emphasizes the importance of appropriate interpretation and reporting of subgroup analyses, including formal assessment of treatment-by-subgroup interactions. A recent systematic review of subgroup analyses in stepped-wedge cluster randomized trials [14] similarly found that subgroup analyses were common, but treatment-by-subgroup interaction results were reported inconsistently.

In crossover studies, subgroup heterogeneity introduces an additional inferential problem because treatment history may itself have subgroup-specific effects. A residual effect of treatment received in an earlier period may differ across patient subgroups even when an overall carryover effect is present. When the scientific question concerns this heterogeneity, the target estimand is a carryover-by-subgroup interaction rather than a population-average carryover effect. A mean model that explicitly includes subgroup effects and the relevant treatment and carryover interactions can directly estimate this heterogeneity. In contrast, a reduced model that omits these subgroup terms estimates a collapsed carryover contrast that does not characterize subgroup-specific carryover heterogeneity. Thus, mean-model specification determines whether the fitted model targets the scientific estimand of interest.

Valid inference also depends on appropriate representation of the correlation induced by repeated measurements within participants. Mean and covariance specifications therefore address distinct but complementary aspects of the analysis: The scientific question defines the estimand, whereas the fitted mean structure determines the model-based target of estimation and whether it aligns with that estimand., whereas the covariance structure affects the uncertainty and efficiency with which that quantity is estimated. Research on mixed models for repeated measures (MMRM) has shown that covariance and testing specifications can affect standard errors, confidence interval coverage, type I error, and power, particularly in smaller or unbalanced studies [15]. However, the implications of these modeling choices have received comparatively little attention when the target estimand is subgroup-specific carryover heterogeneity. In particular, The finite-sample operating characteristics of carryover-by-subgroup interaction inference under estimated covariance, alternative working covariance specifications, and unequal sequence or subgroup allocation have received comparatively limited evaluation within a common crossover framework.

We therefore conducted a simulation-based evaluation of subgroup-specific carryover inference in two-treatment, two-sequence settings. Using simulation, we compared full mean models that explicitly represented subgroup-specific treatment and carryover effects with reduced models that omitted these terms and evaluated analysis of variance and MMRM approaches under alternative covariance specifications. We examined their consequences for estimation of the carryover-by-subgroup interaction, confidence interval coverage, type I error, and power. We further illustrated the framework using observational crossover-like medication sequences from the National Alzheimer's Coordinating Center (NACC) data. By separating specification of the scientific estimand from specification of the within-participant covariance structure, this study aims to provide practical guidance for evaluating heterogeneous carryover effects in crossover and treatment-transition studies. The primary objective is to characterize the finite-sample operating properties of the carryover-by-subgroup interaction test when within participant covariance parameters are estimated and the working covariance structure may be misspecified. The reduced model is included to represent a population-average analysis question, not as an alternative estimator of subgroup heterogeneity. By evaluating the finite-sample performance of subgroup-specific carryover inference across different sample sizes, allocation patterns, and covariance structures, this study provides practical evidence to inform statistical model specification in pharmaceutical research and clinical trials. These findings may help pharmaceutical statisticians recognize potential challenges in analyzing treatment carryover effects and select analysis strategies that better align with their subgroup-specific clinical research objectives.

## 2 | Statistical Methods

We considered a standard two-treatment, two-sequence, two-period crossover design with continuous outcomes measured at baseline and at the end of each treatment period, yielding three repeated measurements per participant. Participants were assigned to one of two treatment sequences, differing only in the order in which treatments A and B were administered. To evaluate the impact of model specification on subgroup-specific carryover inference, we considered 3 analysis approaches: an independence-based linear model fitted by ordinary least squares, referred to as ANOVA for consistency with common crossover terminology, a mixed model for repeated measures with a compound-symmetry covariance structure (MMRM-CS), and a mixed model for repeated measures with an unstructured covariance structure (MMRM-UN). MMRM-CS and MMRM-UN models were estimated by restricted maximum likelihood (REML). Fixed-effect inference used the Kenward-Roger small-sample adjustment, and two-sided 95% Wald-type confidence intervals and tests were based on the corresponding $t$ distribution. The ANOVA comparator was fitted by ordinary least squares.

### 2.1 | Covariance Structure

The 3 analysis approaches differed in their assumptions regarding the covariance among the 3 repeated measurements within each participant. Let

$$\boldsymbol{\epsilon}_j = \left(\epsilon_{j0}, \epsilon_{j1}, \epsilon_{j2}\right)^T$$

denote the residual vector for participant (*j*), corresponding to the baseline, Period 1, and Period 2 measurements, respectively.

For the ANOVA-based linear model, residual errors were assumed to be independent with a common variance,

$$\boldsymbol{\epsilon}_j \sim N(0, \Sigma_{\text{IND}}), \quad \Sigma_{\text{IND}} = \sigma^2 I_3 = \begin{pmatrix} \sigma^2 & 0 & 0 \\ 0 & \sigma^2 & 0 \\ 0 & 0 & \sigma^2 \end{pmatrix} \tag{1}$$

Thus, the ANOVA-based model assumed equal residual variances across the 3 measurements and 0 within-participant residual correlations. The ANOVA comparator was implemented as an ordinary linear model assuming independent, homoscedastic residual errors, it is referred to hereafter as ANOVA for brevity.

For MMRM-CS, the residual vector was assumed to follow a multivariate normal distribution with a compound-symmetry covariance matrix,

$$\boldsymbol{\epsilon}_j \sim N(0, \Sigma_{CS}),$$

$$\Sigma_{CS} = \begin{pmatrix} \sigma^2 & \rho\sigma^2 & \rho\sigma^2 \\ \rho\sigma^2 & \sigma^2 & \rho\sigma^2 \\ \rho\sigma^2 & \rho\sigma^2 & \sigma^2 \end{pmatrix} \tag{2}$$

Under the CS structure, the three repeated measurements have a common residual variance and a common pairwise correlation ($\rho$).

For MMRM-UN, an unstructured covariance matrix was specified,

$$\boldsymbol{\epsilon}_j \sim N(0, \Sigma_{UN}),$$

$$\Sigma_{UN} = \begin{pmatrix} \sigma_1^2 & \sigma_{12} & \sigma_{13} \\ \sigma_{12} & \sigma_2^2 & \sigma_{23} \\ \sigma_{13} & \sigma_{23} & \sigma_3^2 \end{pmatrix} \tag{3}$$

The UN structure allows the three repeated measurements to have distinct residual variances and distinct pairwise covariances, without imposing the equality constraints of the CS structure. Therefore, MMRM-UN provides a more flexible representation of within-participant covariance than MMRM-CS.

**2.2 | Full and Reduced Models**

For each of the 3 analysis approaches described in **Section 2.1**, we fitted a full model incorporating subgroup-specific treatment and carryover effects and a reduced model omitting subgroup-related terms. The full mean model was specified as

$$Y_{jk} = \beta_0 + \beta_b B_{jk} + \beta_p P_{jk} + \beta_t T_{jk} + \beta_c C_{jk} + \beta_g G_j + \beta_{t,g}(G_j \times T_{jk}) + \beta_{c,g}(G_j \times C_{jk}) + \varepsilon_{jk} \tag{4}$$

Here, $Y_{jk}$ denotes the continuous outcome for participant $j$ at measurement $k$, where $k \in \{0,1,2\}$ corresponds to baseline, Period 1, and Period 2, respectively. $B_{jk}$ is the baseline-visit indicator, with $B_{jk} = 1$ for the baseline measurement and 0 otherwise; $P_{jk}$ is the period indicator, with $P_{jk} = 1$ for Period 2 and 0 otherwise; $T_{jk}$ is the treatment indicator, with $T_{jk} = 1$ for active treatment and 0 for placebo; $C_{jk}$ is the carryover indicator, with $C_{jk} = 1$ when participant $j$ received active treatment in the preceding period and 0 otherwise; and $G_j$ is the subgroup indicator, with $G_j = 1$ for the comparison subgroup and $G_j = 0$ for the reference subgroup.

The coefficient $\beta_0$ denotes the intercept; the subscript $b$ in $\beta_b$ denotes the baseline-visit effect; the subscript $p$ in $\beta_p$ denotes the Period 2 effect; the subscript $t$ in $\beta_t$ denotes the treatment effect in the reference subgroup $G_j = 0$; the subscript $c$ in $\beta_c$ denotes the carryover effect in the reference subgroup; and the subscript $g$ in $\beta_g$ denotes the subgroup main effect under the reference levels of the visit, treatment, and carryover indicators. The subscripts $t, g$ in $\beta_{t,g}$ denote the treatment-by-subgroup interaction, representing the difference in the treatment effect between $G_j = 1$ and $G_j = 0$, whereas the subscripts $c, g$ in $\beta_{c,g}$ denote the carryover-by-subgroup interaction, representing the difference in the carryover effect between $G_j = 1$ and $G_j = 0$. Thus, $\beta_{c,g}$ quantifies subgroup heterogeneity in the carryover effect and was the primary parameter of interest.

The reduced mean model omitted the subgroup main effect and both subgroup interaction terms and was specified as

$$Y_{jk} = \beta_0^{\text{global}} + \beta_b^{\text{global}} B_{jk} + \beta_p^{\text{global}} P_{jk} + \beta_t^{\text{global}} T_{jk} + \beta_c^{\text{global}} C_{jk} + \epsilon_{jk}. \tag{5}$$

In the reduced model, the superscript "global" indicates that the corresponding parameter is estimated without conditioning on subgroup membership. Specifically, $\beta_0^{\text{global}}$ denotes the global intercept; the subscript $b$ in $\beta_b^{\text{global}}$ denotes the global baseline-visit effect; the subscript $p$ in $\beta_p^{\text{global}}$ denotes the global Period 2 effect; the subscript $t$ in $\beta_t^{\text{global}}$ denotes the global treatment effect; and the subscript $c$ in $\beta_c^{\text{global}}$ denotes the global carryover effect. Because $G_j$, $G_j \times T_{jk}$, and $G_j \times C_{jk}$ are omitted, the reduced model does not estimate subgroup-specific treatment or carryover effects. In particular, $\beta_c^{\text{global}}$ represents a single pooled carryover contrast across participants rather than the carryover effect within either subgroup or the difference in carryover effects between subgroups.

For subgroup $G_j$, let $\Delta_C(G_j)$ denote the model-based carryover contrast between observations with and without prior-treatment carryover, conditional on the remaining terms in the mean model. The subgroup-specific carryover heterogeneity estimand is

$$\theta_C = \Delta_C(1) - \Delta_C(0).$$

Under the binary 0/1 coding used here, $\Delta_C(0) = \beta_C$, $\Delta_C(1) = \beta_C + \beta_{C\times G}$, and therefore $\theta_C = \beta_{C\times G}$.

Combining the two mean-model specifications with the 3 covariance approaches defined in **Section 2.1** yielded 6 fitted models. The full model fitted under the independence assumption was denoted ANOVA-Full, and the corresponding reduced model was denoted ANOVA-Reduced. The full and reduced models fitted using the CS covariance structure were denoted MMRM-CS-Full and MMRM-CS-Reduced, respectively. Similarly, the full and reduced models fitted using the unstructured covariance specification were denoted MMRM-UN-Full and MMRM-UN-Reduced, respectively. These 6 model names are used throughout the remainder of the manuscript. The full and reduced mean models therefore answer different scientific questions. The full model targets subgroup-specific carryover contrasts and their difference, whereas the reduced model targets a single population-average carryover contrast. The reduced model was included as a benchmark for this alternative analysis question and was not expected to estimate $\beta_{c,g}$.

## 3 | Simulation Study

### 3.1 | Simulation setting

Each simulated dataset represented a three-observation, two-treatment, two-sequence, two-period crossover study based on the model specification and notation introduced in Section 2. The three observations corresponded to baseline, Period 1, and Period 2, with the two treatment sequences administered in opposite orders. Under the simulation coding, Sequence 1 received active treatment before placebo; consequently, only its Period 2 measurement could carry carryover from active treatment. Some large-sample point-estimation properties of the fitted linear models can be characterized analytically when the design matrix and within-participant covariance are treated as fixed and known. In the implemented MMRM analyses, however, covariance parameters were estimated from each dataset by REML, and fixed-effect inference used the Kenward–Roger adjustment. Simulation was therefore used to evaluate the finite-sample operating characteristics of the fitted procedures under covariance estimation, covariance misspecification, and allocation imbalance.

The performance of the fitted models was evaluated using bias, 95% confidence interval coverage, Type I error, and power. These measures were selected following the simulation framework of Morris et al. [16], in which bias assesses systematic deviation of the estimator from the true parameter, coverage assesses the calibration of confidence intervals, and rejection rates under the null and alternative hypotheses quantify empirical Type I error and power, respectively. We used 10,000 independent simulation replicates; for a true Type I error rate of 0.05, the corresponding Monte Carlo standard error was approximately 0.0022. For the full models, Type I error and power referred to the test of $H_0: \beta_{c,g} = 0$. For the reduced models, rejection probabilities referred to the distinct global null $H_0: \beta_c^{global} = 0$. Bias and coverage for the reduced coefficient relative to the reference-subgroup carryover effect were treated as diagnostic measures of estimand mismatch, not as conventional bias or coverage for the reduced model's own target.

To evaluate operating characteristics under design imbalance, we simulated data across 9 distinct allocation patterns. These patterns correspond to the cross-classification of 3 sequence allocation ratios (1:1, 8:7, and 9:7) and 3 subgroup prevalence ratios (1:1, 3:5, and 5:3). These patterns were prespecified to represent balanced and moderately unbalanced treatment-sequence allocations and subgroup prevalences reflecting plausible variation in real-world applications. The operating characteristics of subgroup-by-carryover inference under such allocation imbalance have received limited methodological attention in crossover settings, although subgroup prevalence is known to influence statistical precision and power in treatment-effect heterogeneity analyses [7, 17]. A minimal design-matrix representation of the sequence-by-subgroup combinations is provided in **Supporting Information 1: Table S1.1.**

Simulations were conducted across a range of total sample sizes ($N$ = 10, 15, 20, 25, 30, 40, 50, 60, 80, 100, 130, 160, and 200). Within-participant residual vectors were generated according to the CS and UN covariance structures defined in **Equations (2)** and **(3)**, respectively, and were combined with the fixed-effect mean structure defined in **Section 2.2**. The fixed parameters were set to $\beta_0$ = 20.5, $\beta_b$ = 0.4, $\beta_p$ = -2 and $\beta_g$ = 0. The treatment and carryover effect parameters are summarized in **Table 1**. Scenario labels follow the convention (C_x_T_y), where (x) denotes the prespecified carryover-by-subgroup interaction $\beta_{c,g}$, and (y) denotes the prespecified treatment-by-subgroup interaction $\beta_{t,g}$. These interaction parameters represent the between-subgroup differences in carryover and treatment effects, respectively. Type I error was evaluated under the interaction null $\Delta_C = 0$, including settings in which the two subgroups shared the same nonzero carryover effect, as well as the stronger null in which carryover effects were zero in both subgroups. This distinction allowed calibration of the interaction test to be assessed independently of the presence of an overall carryover signal.

**TABLE 1** Simulation scenarios.

| | | | G = 0 | | G = 1 | |
|---|---|---|---|---|---|---|
| Label | $\beta_{c,g}$ | $\beta_{t,g}$ | $\beta_c$ | $\beta_t$ | $\beta_c + \beta_{c,g}$ | $\beta_t + \beta_{t,g}$ |
| C1.5_T3 | 1.5 | 3 | 1.5 | 3 | 3 | 6 |
| C1.5_T2 | 1.5 | 2 | 1.5 | 3 | 3 | 5 |
| C1.5_T1 | 1.5 | 1 | 1.5 | 3 | 3 | 4 |
| C2_T3 | 2 | 3 | 1 | 3 | 3 | 6 |
| C2_T2 | 2 | 2 | 1 | 3 | 3 | 5 |
| C2_T1 | 2 | 1 | 1 | 3 | 3 | 4 |
| C2.5_T3 | 2.5 | 3 | 1 | 3 | 3.5 | 6 |
| C2.5_T2 | 2.5 | 2 | 1 | 3 | 3.5 | 5 |
| C2.5_T1 | 2.5 | 1 | 1 | 3 | 3.5 | 4 |
| C3_T3 | 3 | 3 | 1 | 3 | 4 | 6 |
| C3_T2 | 3 | 2 | 1 | 3 | 4 | 5 |
| C3_T1 | 3 | 1 | 1 | 3 | 4 | 4 |
| C1.5_T0 | 1.5 | 0 | 1.5 | 3 | 3 | 3 |
| C2_T0 | 2 | 0 | 1 | 3 | 3 | 3 |
| C2.5_T0 | 2.5 | 0 | 0.5 | 3 | 3 | 3 |
| C3_T0 | 3 | 0 | 0 | 3 | 3 | 3 |
| *C0_T3_A | 0 | 3 | 0 | 3 | 0 | 6 |
| *C0_T3_B | 0 | 3 | 1 | 3 | 1 | 6 |
| *C0_T3_C | 0 | 3 | 3 | 3 | 3 | 6 |
| *C00_T00 | 0 | 0 | 0 | 0 | 0 | 0 |

Note: Scenarios marked with an asterisk (*) are used to evaluate Type I error; Scenario labels follow the convention (C_x_T_y), where (x=$\beta_{c,g}$) is the between-subgroup difference in the carryover effect and (y=$\beta_{t,g}$) is the between-subgroup difference in the treatment effect. ($G$=0) is the reference subgroup; the corresponding effects for ($G$=1) are ($\beta_c + \beta_{c,g}$) and ($\beta_t + \beta_{t,g}$).

The covariance matrices $\Sigma_{CS}$ and $\Sigma_{UN}$, used to generate the simulated data under **equations (2) and (3)**, respectively, are provided in **Supporting Information 1: Table S1.2-S1.3**. The covariance settings were selected to span variance and correlation patterns broadly consistent with continuous longitudinal outcomes encountered in practice. For the CS structures, the common marginal variance $\sigma^2$ ranged from 3.0 to 24.0, with within-participant correlations $\rho$ fixed at 0.4 or 0.6. For the UN structures, marginal variances varied across measurements, ranging from 3.0 to 30.0, with pairwise correlations ranging from 0.1 to 0.8. The selected configurations included both heterogeneous correlation patterns and patterns with stronger correlations between adjacent than more distant measurements, thereby representing a range of plausible within-participant covariance structures [5, 18].

Across the simulation scenarios, we varied the total sample size, within-subject covariance structure, the treatment-by-subgroup interaction, and the carryover-by-subgroup interaction in data generation. Each generated dataset was analyzed using the full and reduced ANOVA models and the corresponding MMRM models with CS and UN covariance specifications. The remaining fixed-effect parameters were held constant, and the same subgroup allocation ratio was applied within each sequence. This design allowed the effects of omitting subgroup interactions to be distinguished from those of covariance-structure misspecification. The primary inferential target was the carryover-by-subgroup interaction, while the treatment-by-subgroup interaction was varied to assess whether concurrent treatment heterogeneity affected carryover inference.

All prespecified sample sizes and fitted covariance specifications were evaluated for each scenario. Simulation studies were conducted using R version 4.5.3.

### 3.2 | Simulation Results

The main text presents results under a representative CS data-generating condition, $\Sigma_{CS,15,0.4}$, with a common marginal variance of $\sigma^2 = 15$ and a common within-participant correlation of $\rho = 0.40$, corresponding to a common covariance of 6.0 (**Supporting Information 2: Table S2.1-S2.4**). This condition was selected to illustrate performance under moderate within-participant correlation and an intermediate marginal variance within the simulated CS parameter range. For the full model, performance was evaluated for the carryover-by-subgroup interaction, $\Delta_C$. For the reduced model, performance was evaluated for its fitted carryover coefficient, which represents a collapsed carryover contrast rather than $\Delta_C$.

#### 3.2.1 Performance under the full interaction model

Empirical power for testing $H_0: \Delta_C = 0$ increased with total sample size and the magnitude of the true carryover heterogeneity, as shown in **Figure 1**. At $N = 60$, power for $\Delta_C = 3.0$ was approximately 0.42–0.60 across the analysis and allocation configurations displayed in the **Figure 1**. At $N = 200$, power was approximately 0.35-0.55 for $\Delta_C = 1.5$ and 0.91-1.00 for $\Delta_C = 3.0$, with the results for $\Delta_C = 2.0$ and 2.5 falling between these ranges (**Figure 1**). Under this selected condition, the CS and UN approaches generally produced similar power and were more powerful than ANOVA, particularly at moderate sample sizes (**Figure 1**). For a fixed value of $\Delta_C$, the power curves corresponding to $\Delta_T = 0,1,2,$ and 3 were nearly superimposed, indicating that treatment-effect heterogeneity had little influence on the correctly specified full-model test of the carryover-by-subgroup interaction (**Figure 1**).

Under the null condition $\Delta_C = 0$, empirical Type I error was approximately 3.0%-3.8% for ANOVA and 4.5%-5.6% for the CS and UN approaches. Thus, the CS and UN tests were generally close to the nominal 5% level, whereas the ANOVA test was conservative under the selected CS-generated condition (**Supporting Information 2: Table S2.3**).

Coverage of the nominal 95% confidence intervals for $\Delta_C$ remained approximately 94%–96% for the CS and UN approaches across the evaluated effect-size and allocation configurations. ANOVA coverage was more variable, ranging approximately from 89% to 99%, consistent with the consequences of analyzing correlated repeated measurements under an independence assumption (**Supporting Information 2: Table S2.1**).

The full model estimator of $\Delta_C$ remained centered near the true interaction across sample sizes and effect-size configurations. Absolute bias was generally less than 0.04 for all 3 analysis approaches (**Supporting Information 2: Table S2.1**). These results indicate that inclusion of the subgroup main effect and the treatment-by-subgroup and carryover-by-subgroup interactions allowed direct estimation of the intended subgroup-specific carryover contrast. Covariance specification primarily affected inferential efficiency and confidence-interval calibration.

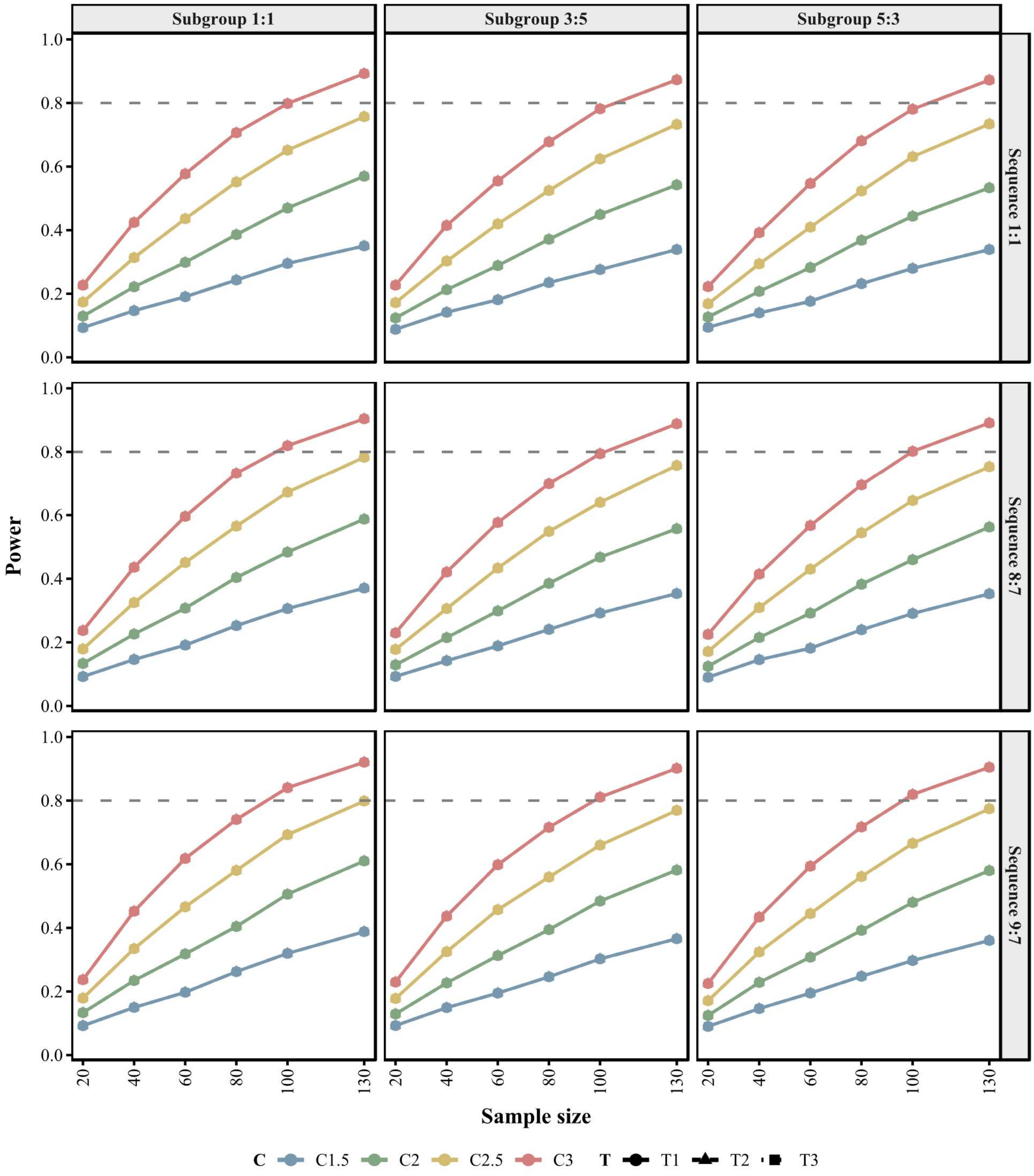

**FIGURE 1** Power for the full model under the selected CS data-generating condition. Power was defined as $\Pr(\text{ reject } H_0\colon \theta = 0 \mid \theta \neq 0\,)$ and estimated by $\widehat{\text{Power}} = \frac{1}{M}\sum_{m=1}^{M} I(p_m \leq \alpha)$, where $M$ is the number of simulation replicates, $p_m$ is the p-value from replicate $m$, and $\alpha = 0.05$. The generic parameter $\theta$ denotes $\beta_{c,g}$ for the interaction-preserving models.

### 3.2.2 Impact of covariance specification

To examine the effect of covariance misspecification separately from misspecification of the subgroup mean structure, the full interaction model was additionally evaluated under the representative unstructured data-generating covariance $\Sigma_{\mathrm{UN},8}$, with marginal variances of 9, 22, and 28 and correlations of 0.70, 0.30, and 0.70 between the three repeated observations (**Supporting Information 1: Table S1.3**). This setting represents heterogeneous within-participant variability and correlation and was selected as an empirically motivated UN condition broadly consistent with the unequal covariance pattern observed in the many real world application. Under this data-generating condition, the UN analysis correctly specified the covariance structure, whereas the CS analysis imposed a common marginal variance and common within-participant correlation; ANOVA did not model within-participant correlation.

**Figure 2** illustrates inferential calibration under the balanced sequence and subgroup allocation setting, with both allocation ratios equal to 1:1. Under the null condition $\Delta_C = 0$, empirical Type I error remained close to the nominal 5% level under the correctly specified UN model across the displayed sample sizes, whereas the misspecified CS model showed persistent inflation to approximately 12%. ANOVA also showed inflation, although less pronounced than that under CS (**Figure 2A**). A corresponding pattern was observed for confidence-interval coverage. Under the representative alternative configuration $\Delta_C = 3.0$ and $\Delta_T = 0$, UN coverage remained close to the nominal 95% level, whereas CS coverage was consistently below 90%; ANOVA coverage was intermediate between the two (**Figure 2B**). Thus, even when the subgroup mean structure was correctly specified, misspecification of the within-participant covariance materially affected the calibration of hypothesis tests and confidence intervals.

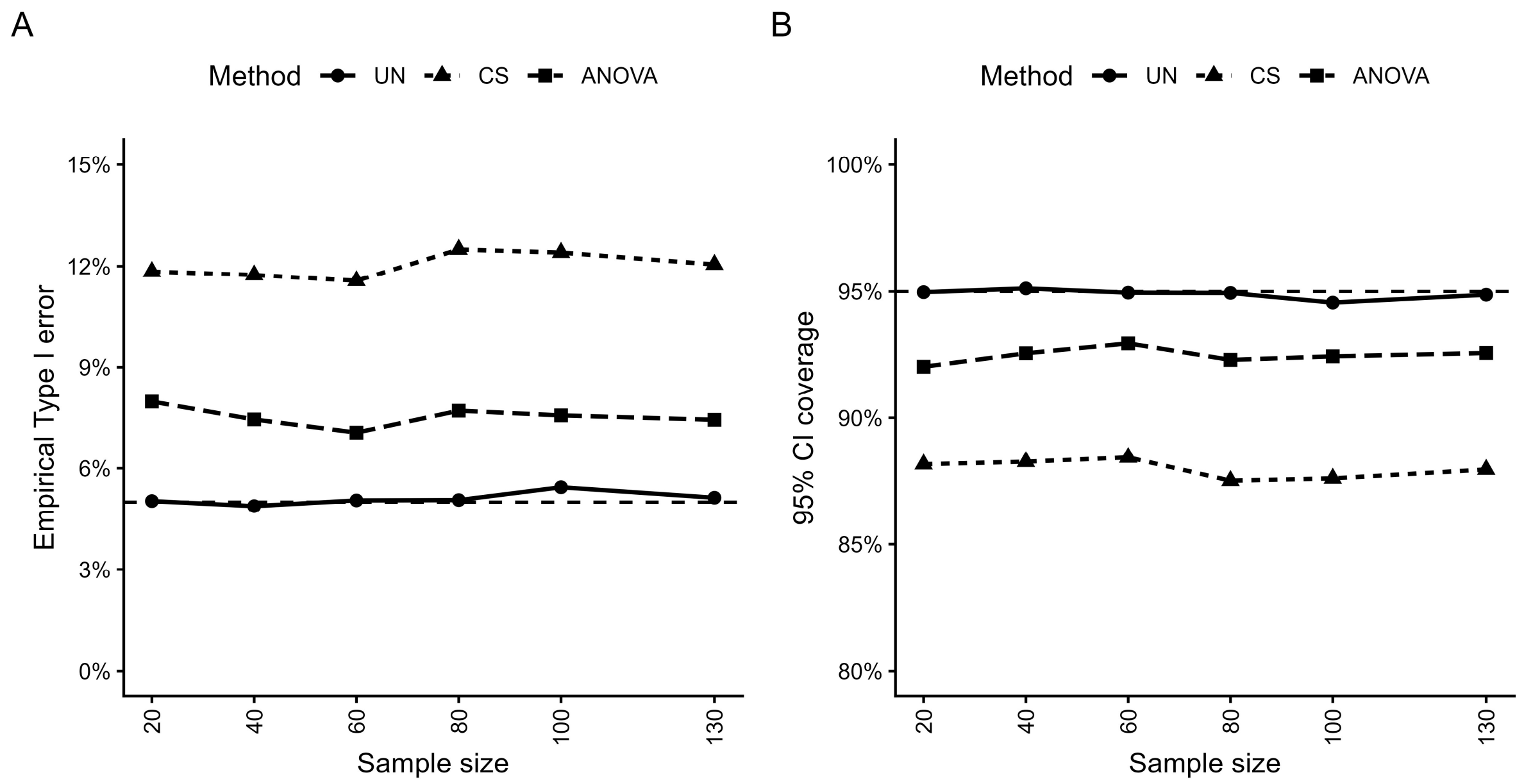

**FIGURE 2** Inferential calibration under the representative unstructured data-generating covariance. $\Sigma_{UN,8}$ with balanced sequence and subgroup allocation. (A) Empirical Type I error for the carryover-by-subgroup interaction under the null condition $\Delta_C = 0$ (C00_T00). (B) Empirical coverage of the nominal 95% confidence interval for the carryover-by-subgroup interaction under $\Delta_C = 3.0$ and $\Delta_T = 0$(C3_T0). Sequence and subgroup allocation ratios were both 1:1. UN represents the correctly specified covariance model, CS represents a misspecified compound-symmetry covariance model, and ANOVA does not model within-participant correlation. Horizontal reference lines indicate the nominal Type I error rate of 0.05 and nominal coverage probability of 0.95, respectively.

The same qualitative pattern was observed across the broader set of evaluated sample sizes, effect-size configurations, and sequence and subgroup allocation ratios. Across these settings, coverage of the nominal 95% confidence interval for $\Delta_C$ was approximately 93.5%-95.6% under the correctly specified UN model, compared with 87.3%-88.7% under the misspecified CS model and 91.4%-93.4% under ANOVA (**Supporting Information 2: Table S2.1**). Under the null condition, empirical Type I error was approximately 4.4%-6.5% for UN, 11.3%-12.7% for CS, and 6.6%-8.6% for ANOVA (**Supporting Information 2: Table S2.3**). In contrast, point estimation was comparatively insensitive to covariance specification: absolute bias in the carryover-by-subgroup interaction estimate remained below 0.04 for both the CS and UN approaches and below 0.06 for ANOVA across the evaluated configurations (**Supporting Information 2: Table S2.1**).

Under alternative hypotheses, the misspecified CS analysis frequently produced higher rejection rates than the correctly specified UN analysis. For example, when $\Delta_C = 3.0$, empirical rejection rates at $N = 60$ were approximately 0.53-0.59 under CS, compared with 0.37-0.44 under UN and 0.35-0.40 under ANOVA; at $N = 200$, the corresponding ranges were approximately 0.93-0.96, 0.87-0.91, and 0.79-0.84, respectively (**Supporting Information 2: Table S2.1**). These higher rejection rates under CS should not be interpreted as a genuine gain in power or efficiency because they occurred together with substantial Type I error inflation and undercoverage. Rather, they reflect anti-conservative inference associated with fitting CS under this heterogeneous UN-generating condition.

Taken together with the CS-generated results in **Section 3.2.1**, these findings distinguish flexibility from misspecification. When the data were generated under CS, fitting the more flexible UN structure produced broadly similar inference, whereas imposing CS on data generated from a heterogeneous UN covariance produced substantial undercoverage and Type I error inflation. The working covariance specification did not change the model-based estimand and had comparatively little influence on point-estimate bias in these scenarios, but it materially affected the calibration of standard errors, confidence intervals, and hypothesis tests.

### 3.2.3 Population-average inference under the reduced model

The reduced model omitted the subgroup main effect and the subgroup-by-treatment and subgroup-by-carryover interaction terms. Consequently, its fitted carryover coefficient, $\hat{\beta}_C^{global}$, represented a collapsed carryover contrast rather than the carryover-by-subgroup interaction $\Delta_C$.

The empirical rejection probability for the reduced-model global carryover test increased with sample size but depended on the absolute subgroup-specific carryover effects and their allocation proportions rather than on $\Delta_C$ alone. At $N = 60$, power ranged approximately from 0.09 to 0.55 across the alternative configurations and from 0.26 to 0.95 at $N = 200$ (**Figure 3; Supporting Information 2: Table S2.2**). Accordingly, these rejection probabilities quantify detection of a nonzero collapsed carryover coefficient and should not be interpreted as power for detecting carryover heterogeneity.

In contrast to the full interaction model, the reduced-model power curves also varied with treatment-effect heterogeneity (**Figure 3**). Thus, after omission of the subgroup-related terms, inference for $\hat{\beta}_c^{global}$ was no longer determined by the subgroup-specific carryover effects alone. When the true carryover effects were zero in both subgroups, empirical Type I error was approximately 4.0%-5.6% for the CS and UN approaches but 8.5%-11% for ANOVA, indicating reasonable calibration for the reduced model's own global null under the repeated-measures approaches but anti-conservative inference under independence (**Supporting Information 2: Table S2.4**).

When $\hat{\beta}_c^{global}$ was evaluated diagnostically against the true carryover effect in the reference subgroup G=0, the estimand mismatch produced persistent bias and progressively poorer coverage. Across the heterogeneous-carryover configurations, diagnostic coverage declined from approximately 76%-94% at $N = 60$ to 36%-92% at $N = 200$, with the greatest undercoverage occurring under larger carryover heterogeneity and imbalanced subgroup allocation. Diagnostic bias increased with $\Delta_C$, ranging approximately from 0.50-1.1 when $\Delta_C = 1.5$ to 1.0-2.0 when $\Delta_C = 3.0$ (**Supporting Information 2: Table S2.2**). Under balanced subgroup allocation, the bias was approximately one-half of $\Delta_C$, consistent with the reduced coefficient averaging information across the two subgroup-specific carryover effects. Thus, increasing sample size narrowed confidence intervals around the collapsed reduced-model target without resolving its discrepancy from the subgroup-specific estimand.

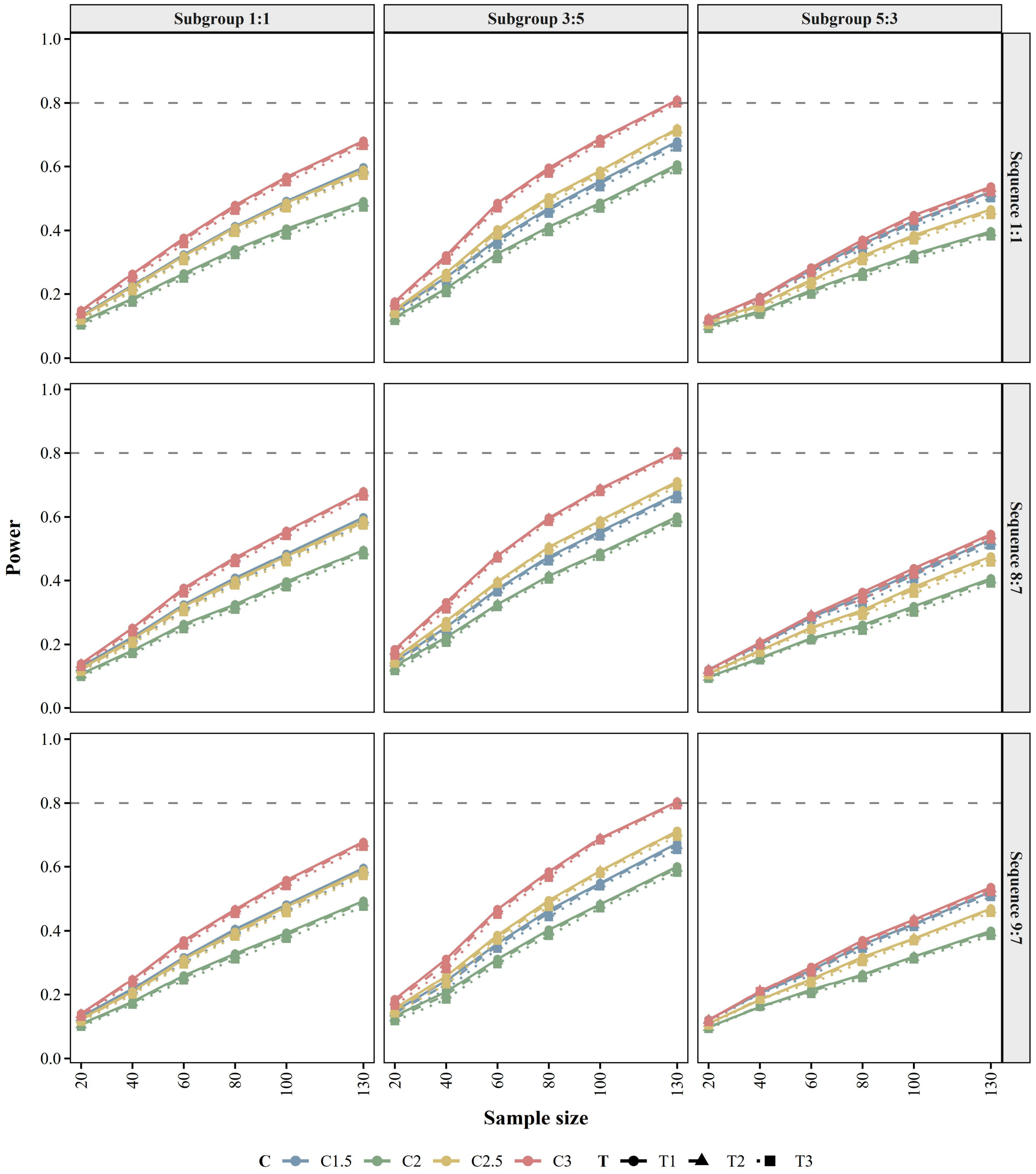

**FIGURE 3** Rejection probability for the reduced-model global carryover test. Power was defined as $\Pr(\text{ reject } H_0\colon \theta = 0 \mid \theta \neq 0\,)$ and estimated by $\widehat{\text{Power}} = \frac{1}{M}\sum_{m=1}^{M} I(p_m \leq \alpha)$, where $M$ is the number of simulation replicates, $p_m$ is the p-value from replicate $m$, and $\alpha$ = 0.05. The generic parameter $\theta$ denotes $\beta_c^{global}$ for the reduced models. Here, $\beta_c^{global}$ denotes the collapsed carryover parameter estimated under the reduced model.

## 4 | Observational treatment-transition application in AD subgroups

### 4.1 | Cohort Curation and Characteristics

To illustrate the modeling distinction using real-world observational data, we constructed a two-period treatment-transition cohort from the NACC medication records. This application was not intended to emulate a randomized crossover efficacy trial and did not assume that the progressive AD disease process was stable or reversible across periods. For this application, “carryover” denotes the model-based association between the preceding regimen and subsequent CDR-SB under the specified mean model, it does not identify a pharmacologic persistence effect. AD treatment switching provides a clinically relevant setting for evaluating subgroup differences in post-switch associations with prior treatment exposure because formal washout periods are generally impractical in progressive cognitive decline. The analyses examined whether the modeled carryover following treatment switching differed across these subgroups and whether omission of subgroup terms, including the carryover-by-subgroup interaction, altered the resulting inference. Accordingly, “carryover” in the NACC application refers to the modeled post-switch association with the preceding regimen and is not intended to represent a directly identified pharmacologic persistence effect.

Recent real-world studies have examined associations of donepezil and memantine, alone or in combination, with survival and longitudinal cognitive trajectories [19, 20]. However, these studies have primarily focused on population-average treatment associations rather than whether the modeled residual effect of a preceding regimen after switching differs across clinically relevant subgroups. Memantine monotherapy was designated treatment A and donepezil plus memantine combination therapy treatment B. Participants in sequence AB received memantine monotherapy in Period 1 followed by combination therapy in Period 2, whereas those in sequence BA received combination therapy in Period 1 followed by memantine monotherapy in Period 2. For both sequences, baseline was defined as an eligible visit at which no AD medication was recorded before initiation of the Period 1 regimen. The Period 1 assessment occurred more than 36 weeks after baseline; this interval was used as a pragmatic cohort-construction criterion rather than as evidence of treatment stabilization or washout. The Period 2 assessment occurred more than 6 weeks later following the observed treatment switch (**Figure 4A**). The 6-week minimum was selected as a pragmatic post-switch assessment interval, rather than as a pharmacologic washout period, with the DOMINO-AD trial providing clinical precedent for assessment 6 weeks after treatment changes involving donepezil and memantine [21].

Three subgroup analyses were conducted separately according to lifetime cigarette-smoking history, behavioral-symptom onset, and APOE ε4 allele status. The smoking history subgroup contrasted participants with versus without a history of smoking more than 100 cigarettes over their lifetime. Smoking is an established modifiable risk factor for dementia, whereas evidence that smoking modifies response to symptomatic AD therapy remains limited [22, 23]. The behavioral subgroup contrasted participants with no behavioral symptoms with those whose behavioral symptoms had a gradual onset. Other onset categories were excluded from this binary subgroup analysis because of insufficient sample sizes for reliable estimation. Behavioral and psychological symptoms are clinically important in AD and may respond to donepezil or memantine, but their role as modifiers of treatment-transition effects has been little studied [24, 25]. The genetic subgroup compared participants with no APOE ε4 allele with those carrying exactly one ε4 allele; participants with two ε4 alleles were excluded. APOE ε4 is a major genetic modifier of sporadic AD risk, and pharmacogenetic studies have suggested possible heterogeneity in symptomatic treatment response, although findings remain inconsistent [26, 27]. Participants with unknown or unavailable subgroup values were excluded from the corresponding subgroup analyses. These 3 NACC subgroup analyses were exploratory and no adjustment for multiple comparisons was applied. Confidence intervals and nominal p-values are interpreted descriptively.

The cohort construction process is shown in **Figure 4B**. Among 54,631 participants in the NACC medication dataset, 49,416 had at least one valid visit date. A total of 499 participants had an observed medication switch between memantine monotherapy and donepezil plus memantine combination therapy, including 204 Sequence AB participants and 295 Sequence BA participants. Application of the baseline medication and visit-timing criteria yielded a final treatment-transition cohort of 94 participants, each contributing three observations corresponding to baseline, Period 1, and Period 2, for a total of 282 observations.

The continuous outcome was the Sum of Boxes score from the CDR® Dementia Staging Instrument (CDR-SB; range, 0-18), selected because of its established longitudinal psychometric properties and clinical interpretability in AD [28]. Mean CDR-SB increased from 3.39 at baseline to 6.22 in Period 1 and 9.47 in Period 2 (**Supporting Information 1: Figure S1**), a pattern consistent with substantial longitudinal clinical worsening over the selected visits.. Although the period term can represent a common temporal shift, it cannot by itself distinguish residual treatment-associated effects from individual disease progression or treatment-selection processes. Moreover, the common Period 2 term represents an average temporal shift and does not account for subgroup-specific disease trajectories.

Because subgroup information was not uniformly available, the analytic sample differed across the 3 subgroup analyses. The behavioral analysis included 91 participants, with 36 classified as having no behavioral symptoms (G=0) and 55 as having a recorded gradual onset of behavioral symptoms (G=1) (**Supporting Information 1: Figure S2**). The APOE ε4 allele count analysis included 78 participants, comprising 29 with no APOE ε4 allele (G=0) and 49 with exactly one ε4 allele (G=1) (**Supporting Information 1: Figure S3**). The smoking history analysis included 79 participants, of whom 44 had not smoked more than 100 cigarettes during their lifetime (G=0) and 35 had (G=1) (**Supporting Information 1: Figure S4**).

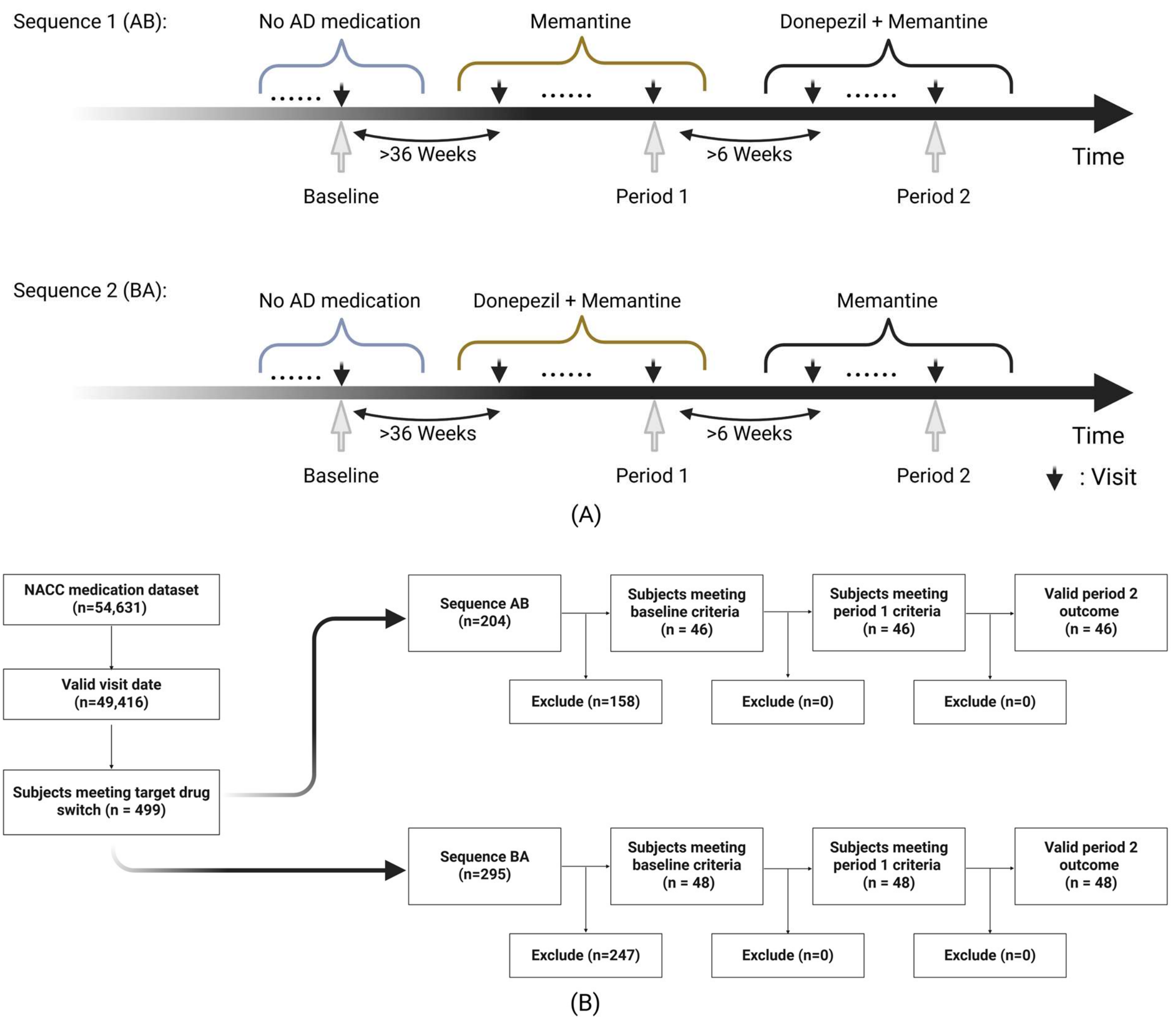

**FIGURE 4** Construction of the NACC Medication-Switch Cohort.

Each subgroup analysis retained observations in all four sequence-by-subgroup cells. For the behavioral subgroup, the corresponding counts were 18 and 28 in AB and 18 and 27 in BA (**Supporting Information 1: Figure S2**). For the APOE ε4 allele count subgroup, the corresponding counts were 12 and 22 in AB and 17 and 27 in BA (**Supporting Information 1: Figure S3**). For smoking history, the AB sequence included 18 participants with G=0 and 19 with G=1, whereas the BA sequence included 26 and 16, respectively (**Supporting Information 1: Figure S4**). Baseline CDR-SB distributions across the four sequence-by-subgroup cells were examined separately for each subgroup to characterize baseline heterogeneity and confirm cell support for model fitting. Such variation is expected in an observational medication-switch cohort and was therefore described rather than interpreted as randomized baseline imbalance.

In the fitted models, baseline CDR-SB was included as the first component of the three-visit repeated outcome vector rather than entered again as a separate baseline covariate. Baseline and post-baseline outcomes were therefore modeled jointly, with their dependence represented through the specified within-subject covariance structure [18]. The repeated-measures treatment-transition design uses within-participant information because each participant contributes observations under both regimens, thereby reducing sensitivity to stable between-participant variability; however, it does not eliminate confounding from time-varying disease severity or treatment-selection processes [29]. Demographic and clinical characteristics of the three subgroup-specific analytic samples are summarized in **Table 2**. P values are descriptive and were not used for covariate selection or causal adjustment.

**TABLE 2** Demographic and clinical characteristics summary.

| Characteristic | Smoking | | | Behavioral | | | Genetic | | |
|---|---|---|---|---|---|---|---|---|---|
| | G=0 (N=44) | G=1 (N=35) | P-value | G=0 (N=36) | G=1 (N=55) | P-value | G=0 (N=29) | G=1 (N=49) | P-value |
| Switch sequence, n (%) | | | 0.2366 | | | 0.9324 | | | 0.762 |
| BA sequence | 26 (59.1%) | 16 (45.7%) | | 18 (50.0%) | 27 (49.1%) | | 17 (58.6%) | 27 (55.1%) | |
| AB sequence | 18 (40.9%) | 19 (54.3%) | | 18 (50.0%) | 28 (50.9%) | | 12 (41.4%) | 22 (44.9%) | |
| Baseline CDR-SB, mean (SD) | 3.93 (3.80) | 2.86 (2.30) | 0.1453 | 2.22 (2.54) | 3.88 (2.89) | 0.0061 | 3.29 (2.87) | 3.24 (2.97) | 0.9442 |
| Age at initial visit, mean (SD) | 72.27 (10.56) | 75.60 (8.02) | 0.127 | 75.14 (9.19) | 72.27 (9.16) | 0.1484 | 75.72 (9.05) | 73.24 (9.53) | 0.2614 |
| Years of education, mean (SD) | 14.27 (3.71) | 15.62 (3.34) | 0.1017 | 15.44 (3.51) | 15.13 (3.77) | 0.6908 | 14.89 (3.83) | 15.16 (3.60) | 0.7575 |
| Sex | | | 0.0462 | | | 0.8045 | | | 0.8566 |
| Female, n (%) | 25 (56.8%) | 12 (34.3%) | | 16 (44.4%) | 23 (41.8%) | | 13 (44.8%) | 23 (46.9%) | |
| Male, n (%) | 19 (43.2%) | 23 (65.7%) | | 20 (55.6%) | 32 (58.2%) | | 16 (55.2%) | 26 (53.1%) | |
| Race | | | 0.5999 | | | 0.5059 | | | 0.3217 |
| Asian, n (%) | 0 | 1 (2.9%) | | 0 | 1 (1.8%) | | 0 | 1 (2.0%) | |
| Black or African American, n (%) | 2 (4.5%) | 3 (8.6%) | | 3 (8.3%) | 2 (3.6%) | | 2 (6.9%) | 2 (4.1%) | |
| Other, n (%) | 3 (6.8%) | 2 (5.7%) | | 1 (2.8%) | 4 (7.3%) | | 3 (10.3%) | 1 (2.0%) | |
| White, n (%) | 39 (88.6%) | 29 (82.9%) | | 32 (88.9%) | 48 (87.3%) | | 24 (82.8%) | 45 (91.8%) | |
| History of stroke | | | 0.8139 | | | 0.63 | | | 0.3641 |
| No, n (%) | 42 (95.5%) | 33 (94.3%) | | 31 (86.1%) | 43 (78.2%) | | 25 (86.2%) | 38 (77.6%) | |
| Yes, n (%) | 2 (4.5%) | 2 (5.7%) | | 1 (2.8%) | 2 (3.6%) | | 0 | 3 (6.1%) | |
| History of diabetes | | | 0.698 | | | 0.0166 | | | 0.9511 |
| No, n (%) | 39 (88.6%) | 30 (85.7%) | | 32 (88.9%) | 36 (65.5%) | | 23 (79.3%) | 38 (77.6%) | |
| Yes, n (%) | 5 (11.4%) | 5 (14.3%) | | 0 | 9 (16.4%) | | 2 (6.9%) | 3 (6.1%) | |
| History of hypertension | | | 0.7827 | | | 0.5954 | | | 0.8649 |
| No, n (%) | 24 (54.5%) | 18 (51.4%) | | 18 (50.0%) | 23 (41.8%) | | 12 (41.4%) | 22 (44.9%) | |
| Yes, n (%) | 20 (45.5%) | 17 (48.6%) | | 14 (38.9%) | 22 (40.0%) | | 13 (44.8%) | 19 (38.8%) | |
| Depression episodes more than two years ago | | | 0.0559 | | | 0.0732 | | | 0.6877 |
| No, n (%) | 38 (86.4%) | 24 (68.6%) | | 29 (80.6%) | 32 (58.2%) | | 21 (72.4%) | 31 (63.3%) | |
| Yes, n (%) | 6 (13.6%) | 11 (31.4%) | | 3 (8.3%) | 13 (23.6%) | | 4 (13.8%) | 10 (20.4%) | |

| Characteristic | Smoking | | | Behavioral | | | Genetic | | |
|---|---|---|---|---|---|---|---|---|---|
| | G=0 (N=44) | G=1 (N=35) | P-value | G=0 (N=36) | G=1 (N=55) | P-value | G=0 (N=29) | G=1 (N=49) | P-value |

Note: For the smoking subgroup, G=0 indicates no history of smoking more than 100 cigarettes during the lifetime and G=1 indicates a history of smoking more than 100 cigarettes. For the behavioral subgroup, G=0 indicates no behavioral symptoms and G=1 indicates behavioral symptoms with a recorded gradual onset. For the genetic subgroup, G=0 indicates no APOE ε4 allele and G=1 indicates exactly one APOE ε4 allele; participants with two ε4 alleles were excluded from this comparison. Participants with unavailable subgroup information were excluded from the corresponding analysis. Continuous variables are summarized as mean (SD), and categorical variables as n (%). Percentages were calculated using the subgroup sample size as the denominator and may not sum to 100% because of missing data.

### 4.2 | Model specification

The NACC application used the full and reduced mean-model structures defined in **Equations (4) and (5)**, respectively. Both models were fitted under the compound-symmetry and unstructured within-subject covariance specifications defined in **Equations (2) and (3)**. Application-specific coding was used for the two active treatment regimens. The subgroup indicator $G_j$ was coded as 0 or 1 according to the prespecified binary comparison for each characteristic. Depending on the specific analysis, $G_j = 0$ indicated no history of smoking more than 100 cigarettes during the lifetime, zero APOE ε4 alleles or no behavioral symptoms, while $G_j = 1$ denoted a history of smoking more than 100 cigarettes during the lifetime, the presence of exactly one APOE ε4 allele or behavioral symptoms with a gradual onset. The current-treatment indicator $T_{jk}$ was coded as 1 for donepezil-plus-memantine therapy and 0 for memantine monotherapy and baseline. The carryover indicator $C_{jk}$ was coded as 1 at Period 2 following prior donepezil-plus-memantine therapy and 0 following prior memantine monotherapy, with $C_{jk} = 0$ at baseline and Period 1. Under this coding, the carryover contrast was $\beta_c$ for $G = 0$ and $\beta_c + \beta_{c,g}$ for $G = 1$. For the NACC application, $\beta_{c,g}$ was interpreted as an associational estimand: the between-subgroup difference in model-based prior-regimen contrasts at Period 2, conditional on the included mean-model terms. It was not interpreted as a causal treatment effect or pharmacologic carryover parameter. In the reduced model, the reported collapsed global carryover contrast was $\beta_c^{global}$.

The full and reduced models were fitted by restricted maximum likelihood using PROC MIXED, with Kenward–Roger denominator degrees of freedom for (t) tests and 95% confidence intervals. For likelihood-ratio tests of nested fixed-effect specifications, the models were refitted using maximum likelihood estimation.

### 4.3 | Application results

The full and reduced models were applied using three participant-level subgroup variables: lifetime cigarette-smoking history, number of APOE ε4 alleles, and mode of onset of behavioral symptoms. The corresponding analyses included 79, 78, and 91 participants, respectively. For cigarette-smoking history, the comparison was between participants who had and had not smoked more than 100 cigarettes during their lifetime. For number of APOE ε4 alleles, the comparison was restricted to participants with zero versus one APOE ε4 allele; participants with two alleles were excluded. For mode of onset of behavioral symptoms, participants with no behavioral symptoms were compared with those whose behavioral symptoms had a gradual onset. Because higher CDR-SB scores indicate greater cognitive and functional impairment, a positive carryover contrast indicates an association of prior donepezil-plus-memantine exposure with higher subsequent CDR-SB relative to prior memantine-only exposure.

Under the CS covariance specification, both the lifetime cigarette-smoking history and number of APOE ε4 alleles analyses indicated larger carryover contrasts in subgroup 1, whereas the direction of subgroup heterogeneity was reversed for mode of onset of behavioral symptoms (**Table 3**). For lifetime cigarette-smoking history, the carryover contrast was small and not statistically significant among participants who had not smoked more than 100 cigarettes during their lifetime (estimate

= 1.34 CDR-SB points; $p = 0.37$) but was larger , with a nominal p value below 0.05 among those who had (estimate = 4.63; $p < 0.01$). The estimated carryover interaction was 3.29 CDR-SB points (95% CI: 0.83, 5.75; $p = 0.01$) (**Table 3**), providing nominal evidence, in this exploratory analysis, of a larger model based prior regimen contrast among participants with this lifetime smoking history. In comparison, the reduced CS model yielded a single global carryover contrast of 2.71 points (95% CI: −0.07, 5.48; $p = 0.06$) (**Table 3**). A similar pattern was observed for number of APOE ε4 alleles. The carryover contrast was not statistically significant among participants with no APOE ε4 allele (estimate = 1.60; $p = 0.32$) but was positive and statistically significant among those with one ε4 allele (estimate = 4.50; $p < 0.01$). The corresponding interaction estimate was 2.90 points (95% CI: 0.48, 5.32; $p = 0.02$), whereas the reduced model produced a global contrast of 3.42 points (95% CI: 0.64, 6.20; $p = 0.02$) (**Table 3**). In contrast, the mode of onset of behavioral symptoms analysis showed a larger carryover contrast among participants with no behavioral symptoms (estimate = 5.05; $p < 0.001$) than among those with gradual onset of behavioral symptoms (estimate = 2.34; $p = 0.08$). The negative interaction estimate of −2.71 points (95% CI: −4.99, −0.43; $p = 0.02$) reflected this reduction in the gradual-onset subgroup. The reduced CS model yielded a statistically significant global contrast of 3.43 points (95% CI: 0.94, 5.92; $p = 0.01$) (**Table 3**) but did not represent the difference between the two subgroups.

Under the UN covariance specification, the directions and relative magnitudes of the subgroup-specific carryover contrasts were generally consistent with the CS results, although the estimates were less precise. The estimated carryover contrast remained small and not statistically significant among participants who had not smoked more than 100 cigarettes (estimate = 1.10; $p = 0.54$) but was positive and statistically significant among those who had (estimate = 4.51; $p = 0.02$). The interaction estimate was 3.41 points (95% CI: 0.23, 6.59; $p = 0.04$), whereas the reduced UN model yielded a nonsignificant global contrast of 2.47 points (95% CI: -0.86, 5.79; p = 0.14) (**Table 3**). The carryover contrast was not statistically significant among participants with no ε4 allele (estimate = 1.46; p = 0.46) but was positive and statistically significant among those with one ε4 allele (estimate = 4.43; p = 0.02). The interaction estimate remained positive at 2.97 points, although its confidence interval included zero (95% CI: −0.07, 6.02; p = 0.06). The reduced UN model similarly produced a positive but statistically nonsignificant global contrast of 3.27 points (95% CI: -0.16, 6.70; p = 0.06) (**Table 3**). For mode of onset of behavioral symptoms, the carryover contrast remained larger among participants with no behavioral symptoms (estimate = 4.73; p = 0.01) than among those with gradual onset (estimate = 2.07; p = 0.21). The interaction estimate was -2.66 points (95% CI: -5.57, 0.25; p = 0.07), indicating the same direction of subgroup heterogeneity as under CS but with greater uncertainty. The reduced UN model yielded a statistically significant global contrast of 3.16 points (95% CI: 0.15, 6.18; p = 0.04) (**Table 3**). Across the three examples, the reduced models provided only a single global carryover coefficient and therefore did not decompose the global prior-regimen association into subgroup-specific contrasts.

**TABLE 3** Carryover estimates from the full and reduced models.

| Subgroup | Model specification | Covariance structure | Estimate label | Estimate | SE | DF | t value | P value | Lower | Upper |
|---|---|---|---|---|---|---|---|---|---|---|
| Lifetime cigarette-smoking history | Full | CS | Carryover G=0 | 1.34 | 1.48 | 228.00 | 0.90 | 0.37 | -1.58 | 4.25 |
| | Full | CS | Carryover G=1 | 4.63 | 1.60 | 227.00 | 2.89 | <0.01 | 1.48 | 7.78 |
| | Full | CS | Carryover interaction | 3.29 | 1.25 | 174.00 | 2.64 | 0.01 | 0.83 | 5.75 |
| | Full | UN | Carryover G=0 | 1.10 | 1.77 | 83.70 | 0.62 | 0.54 | -2.41 | 4.61 |
| | Full | UN | Carryover G=1 | 4.51 | 1.97 | 86.60 | 2.29 | 0.02 | 0.60 | 8.42 |
| | Full | UN | Carryover interaction | 3.41 | 1.60 | 82.20 | 2.13 | 0.04 | 0.23 | 6.59 |
| | Reduced | CS | Carryover | 2.71 | 1.41 | 232.00 | 1.92 | 0.06 | -0.07 | 5.48 |
| | Reduced | UN | Carryover | 2.47 | 1.67 | 77.00 | 1.48 | 0.14 | -0.86 | 5.79 |
| Number of APOE ε4 alleles | Full | CS | Carryover G=0 | 1.60 | 1.60 | 218.00 | 1.00 | 0.32 | -1.55 | 4.75 |
| | Full | CS | Carryover G=1 | 4.50 | 1.47 | 223.00 | 3.05 | <0.01 | 1.59 | 7.40 |
| | Full | CS | Carryover interaction | 2.90 | 1.22 | 167.00 | 2.37 | 0.02 | 0.48 | 5.32 |
| | Full | UN | Carryover G=0 | 1.46 | 1.96 | 94.00 | 0.74 | 0.46 | -2.43 | 5.35 |
| | Full | UN | Carryover G=1 | 4.43 | 1.82 | 85.80 | 2.44 | 0.02 | 0.82 | 8.05 |
| | Full | UN | Carryover interaction | 2.97 | 1.53 | 79.90 | 1.94 | 0.06 | -0.07 | 6.02 |
| | Reduced | CS | Carryover | 3.42 | 1.41 | 228.00 | 2.42 | 0.02 | 0.64 | 6.20 |
| | Reduced | UN | Carryover | 3.27 | 1.72 | 84.50 | 1.90 | 0.06 | -0.16 | 6.70 |
| Mode of onset of behavioral symptoms | Full | CS | Carryover G=0 | 5.05 | 1.43 | 259.00 | 3.54 | <0.001 | 2.24 | 7.86 |
| | Full | CS | Carryover G=1 | 2.34 | 1.34 | 263.00 | 1.75 | 0.08 | -0.30 | 4.98 |
| | Full | CS | Carryover interaction | -2.71 | 1.16 | 201.00 | -2.35 | 0.02 | -4.99 | -0.43 |
| | Full | UN | Carryover G=0 | 4.73 | 1.74 | 105.00 | 2.71 | 0.01 | 1.27 | 8.19 |
| | Full | UN | Carryover G=1 | 2.07 | 1.63 | 103.00 | 1.27 | 0.21 | -1.16 | 5.30 |
| | Full | UN | Carryover interaction | -2.66 | 1.47 | 92.50 | -1.81 | 0.07 | -5.57 | 0.25 |
| | Reduced | CS | Carryover | 3.43 | 1.26 | 268.00 | 2.71 | 0.01 | 0.94 | 5.92 |
| | Reduced | UN | Carryover | 3.16 | 1.52 | 96.40 | 2.08 | 0.04 | 0.15 | 6.18 |

Likelihood-ratio tests were used to compare nested fixed-effects specifications within each covariance structure. Because the compared models differed in their fixed-effects terms, all models used in the LRTs were refitted by maximum likelihood rather than restricted maximum likelihood. For lifetime cigarette-smoking history, removal of the carryover-by-subgroup interaction significantly worsened model fit under both CS (LR = 7.12, $p = 0.01$) and UN (LR = 4.71, $p = 0.03$), whereas the treatment-by-subgroup interaction was not supported. The joint full-versus-reduced comparison did not reach statistical significance under either CS ($p = 0.05$) or UN ($p = 0.10$) (**Supporting Information 1: Table S1.4**). For APOE ε4 allele status, the ML likelihood-ratio comparison yielded nominal evidence for retaining the carryover-by-subgroup interaction under CS (LR = 5.74, $p = 0.02$) and showed borderline evidence under UN (LR = 3.88, $p = 0.05$), whereas the joint 3-df comparisons were not significant (**Supporting Information 1: Table S1.6**). For behavioral-symptom onset, the ML likelihood-ratio comparison yielded nominal evidence for retaining the carryover-by-subgroup interaction under CS (LR = 5.64, $p = 0.02$) but not under UN ($p = 0.06$). However, the joint full-versus-reduced comparison was significant under both CS ($p = 0.04$) and UN ($p = 0.03$), indicating that the subgroup terms collectively improved model fit (**Supporting Information 1: Table S1.8**). Thus, nominal evidence for the smoking-history interaction was observed under both covariance specifications, whereas the APOE and behavioral findings were more sensitive to the covariance structure or inferential procedure. These contrasts describe associations with prior regimen exposure within the selected medication-transition cohort and should not be interpreted as evidence that prior combination therapy caused subsequent worsening in CDR-SB or that any of the evaluated subgroup characteristics is an established treatment modifier.

Covariance-structure fit was summarized from the REML-fitted models using the −2 restricted log likelihood, AIC, AICc, and BIC. Within both the full and reduced fixed-effects specifications, the UN covariance structure produced lower fit criteria than CS for lifetime cigarette-smoking history (**Supporting Information 1: Table S1.5**), APOE ε4 allele status (**Supporting Information 1: Table S1.7**), and behavioral-symptom onset (**Supporting Information 1: Table S1.9**). Because these criteria were obtained under REML, they were used only to compare covariance structures fitted with the same fixed-effects specification; comparisons between the full and reduced fixed-effects models were based on the ML-fitted LRTs reported above.

## 5 | Discussion

We evaluated the finite-sample operating characteristics of subgroup-specific carryover inference under alternative mean and covariance specifications in a two-treatment, two-sequence setting. The primary practical contribution is to distinguish three components that are sometimes conflated: the scientific estimand, the mean structure used to represent the target of estimation, and the covariance specification used for uncertainty quantification. The reduced model was included as a population-average benchmark rather than as a candidate estimator of the carryover-by-subgroup interaction. The simulation results demonstrated that omission of subgroup terms changes the quantity being estimated, whereas covariance misspecification can materially affect uncertainty quantification and hypothesis testing even when the subgroup-specific estimand is retained (**Figures 2 and 3**). Thus, mean-model and covariance specification address related but distinct questions: the former determines whether the model targets the scientific estimand of interest, whereas the latter influences the validity and efficiency of inference for that estimand. Some point-estimation consequences of omitting subgroup interactions, including the interpretation of the reduced coefficient as a weighted projection of subgroup-specific effects, follow from standard linear-model theory. The role of the simulation was therefore not to establish that an omitted interaction cannot be estimated, but to quantify the empirical Type I error, power, confidence-interval coverage, and bias of the implemented OLS and REML/Kenward–Roger procedures when covariance parameters were estimated and the working covariance could be misspecified. The covariance findings are consistent with established MMRM principles; their contribution here is to quantify the magnitude of finite-sample calibration differences for this specific three-visit interaction problem rather than to introduce a new covariance result.

This distinction extends beyond the specific settings evaluated in our simulations. In studies involving repeated observations, treatment sequences, and clinically meaningful subgroups, heterogeneity may arise not only in treatment response but also in the persistence of effects from previous treatment. Consequently, models that adequately represent average treatment, period, or sequence effects may still fail to address the scientific question of interest if subgroup-specific treatment or carryover interactions are omitted. The distinction between mean-model specification and covariance specification may also be relevant in repeated-measures settings with ordered treatment exposures, where associations with prior treatment are scientifically meaningful.

A second contribution of this study is the simulation-based reference resource provided in **Supporting Information 2**. The complete simulation results span prespecified covariance structures, sample sizes, sequence allocations, subgroup allocations, and effect configurations and report bias, confidence-interval coverage, Type I error, and power across alternative analysis strategies (**Supporting Information 2: Tables S2.1-S2.4**). These results extend substantially beyond the representative scenarios presented in the main text (**Figures 1-3**). Rather than serving solely as documentation of the simulation study, this reference grid provides a practical lookup resource for investigators to identify prespecified scenarios that approximate the design and covariance characteristics of a planned or observed study and examine the corresponding operating characteristics. Importantly, these quantities represent repeated-sampling properties under prespecified data-generating conditions; they should not be interpreted as post hoc probabilities that a particular observed result is correct or as a substitute for study-specific simulation when design features differ materially from those represented in the grid.

The NACC application illustrates the practical distinction between subgroup-specific and collapsed prior-regimen associations; it was not used to validate the simulation grid or to conduct post hoc power assessment.. Investigators can first define the subgroup-specific carryover estimand and fit an interaction-preserving repeated-measures model, then characterize the application according to features represented in the simulation grid, including sample size, sequence and subgroup allocation, interaction magnitude, and within-participant covariance pattern. The lifetime smoking-history, APOE ε4 allele-count, and behavioral-symptom analyses provide three empirically motivated examples of this process (**Table 3; Supporting Information 1: Tables S1.4, S1.6, and S1.8**). Across these analyses, carryover-by-subgroup point estimates were similar under CS and UN covariance specifications, whereas the associated uncertainty differed (**Table 3**). Simulation scenarios with comparable design characteristics provided context for interpreting these differences and demonstrated that settings with relatively small bias and approximately nominal confidence-interval coverage could nevertheless differ in their ability to detect subgroup-specific carryover heterogeneity (**Supporting Information 2: Table S2.1**). Because bias, coverage, Type I error, and power cannot be estimated from a single observed dataset, the simulation grid complements the empirical analysis by providing repeated-sampling context without converting the simulation results into post hoc power calculations.

The NACC analysis also demonstrates the broader applicability of the framework beyond randomized crossover trials. Although the NACC medication sequences were observational and should not be interpreted as randomized crossover treatment assignments, they provide an illustrative treatment-transition setting in which repeated outcomes, prior treatment exposure, and clinically meaningful subgroup characteristics coexist. In this context, the framework translates a clinical question about whether residual treatment associations differ across patient subgroups into an explicit interaction-based estimand while accounting for within-participant correlation. The NACC application should therefore be viewed as an illustration of the analytical framework rather than as evidence that the observed medication transitions have causal carryover effects.

Several limitations define the scope of these conclusions. The simulations considered continuous Gaussian outcomes, three fully observed measurements, fixed binary subgroup membership, common covariance structures across sequences and subgroups, and prespecified balanced or moderately unequal allocations (**Table 1; Supporting Information 1: Tables S1.1-S1.3**). Consequently, the current results do not establish operating characteristics for missing or non-Gaussian outcomes, informative discontinuation, severe or sequence-dependent imbalance, subgroup-specific covariance structures, subgroup misclassification, or time-varying subgroup status. The comparator set was intentionally limited to an interaction-preserving model and a population-average benchmark. We did not conduct a comprehensive comparison of all strategies used in practice, such as separate within-subgroup analyses or partially specified models that retain $G$ and $G \times T$ but omit $G \times C$. Conclusions are therefore limited to the candidate analyses evaluated here. The NACC application introduces additional limitations because treatment transitions were observational rather than randomized, no formal washout period was present, subgroup analyses were modest in size and used partially different analytic cohorts, and some sparse subgroup categories required restriction (**Figure 3; Supporting Information 1: Figures S1.1-S1.4 and Tables S1.4-S1.8**). The resulting subgroup contrasts should therefore be interpreted as exploratory associations rather than causal or multiplicity-controlled evidence of differential treatment benefit or harm. Because AD is progressive, and because treatment switching and visit timing were not randomized, the modeled prior-regimen contrasts may also reflect subgroup-specific disease trajectories, treatment escalation, clinician decision-making, or other time-varying confounding. The common period effect cannot fully separate these processes from an association with the preceding regimen.

Future work should extend the framework and accompanying simulation reference to missing and non-Gaussian outcomes, more severe imbalance, subgroup-specific covariance structures, additional periods and sequences, irregular treatment and visit timing, and robust small-sample inference when both mean and covariance structures are uncertain. More broadly, this study provides an estimand-centered approach to subgroup-specific carryover inference by separating 3 components that are sometimes conflated: the scientific question defines the estimand, the mean structure determines whether that estimand is represented by the model, and the covariance structure influences the validity and efficiency of inference. Taken together, the results support a practical distinction between targeting subgroup-specific carryover heterogeneity and estimating a collapsed population-average carryover contrast. Inference for the former requires explicit specification of the interaction of interest, while uncertainty quantification remains sensitive to the working covariance structure in finite samples.

**Data Availability Statement**

The R code used to generate the simulation data, implement the analysis models, summarize the simulation results, and reproduce the simulation figures is available at the project GitHub repository: https://github.com/xiaoqinghuanglab/SubgroupCrossover. The detailed simulation results and the SAS code used for the NACC medication-switch application are also provided in here. No NACC participant-level data or derived individual-level datasets are included in the repository.

The real-world application used data from the June 2025 data freeze of the NACC. Individual-level NACC data cannot be redistributed by the authors because access is governed by NACC data-use policies. Qualified researchers may request access to NACC data through the NACC data request process at https://www.naccdata.org/data-request-process/.

**Acknowledgement**

The authors thank Dr. Zhou Feng, mathematical statistician at the U.S. Food and Drug Administration, for his valuable feedback on this work. The views expressed are those of the authors and do not represent the views of the U.S. Food and Drug Administration.

The NACC database is funded by NIA/NIH Grant U24 AG072122. NACC data are contributed by the NIA-funded ADRCs: P30 AG062429 (PI James Brewer, MD, PhD), P30 AG066468 (PI Oscar Lopez, MD), P30 AG062421 (PI Teresa Gomez-Isla, MD), P30 AG066509 (PI Thomas Grabowski, MD), P30 AG066514 (PI Mary Sano, PhD), P30 AG066530 (PI Helena Chui, MD, Arthur Toga, PhD), P30 AG066507 (PI Marilyn Albert, PhD), P30 AG066444 (PI David Holtzman, MD), P30 AG066518 (PIs Lisa Silbert, MD, Kevin Duff, PhD), P30 AG066512 (PI Thomas Wisniewski, MD), P30 AG066462 (PI Scott Small, MD), P30 AG072979 (PI David Wolk, MD), P30 AG072972 (PIs Charles DeCarli, MD, Rachel Whitmer, PhD), P30 AG072976 (PI Andrew Saykin, PsyD), P30 AG072975 (PI Julie Schneider, MD, MS), P30 AG072978 (PI Ann McKee, MD), P30 AG072977 (PI Robert Vassar, PhD), P30 AG066519 (PI Joshua Grill, PhD), P30 AG062677 (PIs Brad Boeve, MD, Ronald Petersen, MD, PhD), P30 AG079280 (PI Jessica Langbaum, PhD), P30 AG062422 (PI Gil Rabinovici, MD), P30 AG066511 (PI Allan Levey, MD, PhD), P30 AG072946 (PI Linda Van Eldik, PhD), P30 AG062715 (PI Sanjay Asthana, MD, FRCP), P30 AG072973 (PI Russell Swerdlow, MD), P30 AG066506 (PIs Glenn Smith, PhD, ABPP, David Lowenstein, PhD, Ranjan Duara, MD), P30 AG066508 (PIs Stephen Strittmatter, MD, PhD, Christopher Van Dyck, MD), P30 AG066515 (PI Victor Henderson, MD, MS), P30 AG072947 (PI Suzanne Craft, PhD), P30 AG072931 (PI Henry Paulson, MD, PhD), P30 AG066546 (PIs Sudha Seshadri, MD, Gladys Maestre, MD, PhD), P30 AG086401 (PI Erik Roberson, MD, PhD), P30 AG086404 (PI Gary Rosenberg, MD), P30 AG086403 (PI Angela Jefferson, PhD), P30 AG072958 (PIs Heather Whitson, MD, Gwenn Garden, MD, PhD), P30 AG072959 (PI Jagan Pillai, MD, PhD), P30 AG092752 (Ihab Hajjar, MD, MS).

**Funding**

This study is supported by an award to X.H from the Ralph W. and Grace M. Showalter Research Trust and the Indiana University School of Medicine. The content is solely the responsibility of the authors and does not necessarily represent the official views of the Showalter Research Trust or the Indiana University School of Medicine. This research is also supported by the Indiana University Seed Grant to X.H.

*Role of the Funder/Sponsor:* The funding organizations had no role in the design and conduct of the study; collection, management, analysis, and interpretation of the data; preparation, review, or approval of the manuscript; and decision to submit the manuscript for publication.

## Conflicts of Interest

The authors declare no conflicts of interest.